\documentclass[aps]{revtex4}
\def\beq{\begin{eqnarray}}
\def\eeq{\end{eqnarray}}
\begin{document}
\title{\bf Fine-grained state counting for black holes in loop quantum gravity}
\author{A. Ghosh\thanks{amit.ghosh@saha.ac.in}}
\author{P. Mitra\thanks{parthasarathi.mitra@saha.ac.in}}

\affiliation{Saha Institute of Nuclear Physics,\\
1/AF Bidhannagar, Calcutta 700064}
\begin{abstract}
A state of a black hole in loop quantum gravity is given by a
distribution of spins on punctures on the horizon. The distribution
is of the Boltzmann type, with the area playing the r\^{o}le of the
energy. In investigations where the total area was kept
approximately constant, there was a kind of thermal equilibrium
between the spins which have the same analogue temperature
and the entropy was proportional to the area. 
If the area is precisely fixed, however, multiple constraints appear,
different spins have different analogue temperatures and the
entropy is not strictly linear in the area, but is bounded by a
linear rise. 
\end{abstract}
\maketitle
\bigskip 
\bigskip
\section{Introduction}

Black holes are generally regarded as pure manifestations
of gravity. Apart from the obvious geometric properties
of black holes, they have been known to show intriguing
thermodynamic features even though at the classical level
they do not allow anything to escape and cannot have
non-vanishing temperatures. The area of the {\it horizon}
of a black hole was shown to behave like an entropy
in \cite{Bek}. Subsequently the quantum field theory of
a particle in the field of a classical black hole
\cite{Hawk} led to the assignment of a temperature to
a black hole and the area was quantitatively interpreted
as a measure of the thermodynamical entropy.
There have been many attempts to understand this entropy
in a quantum theory of gravity. Recently, the framework 
known as loop quantum gravity has yielded a detailed 
prescription for counting of
microscopic quantum states corresponding to a black hole
\cite{ash}. The quantum states are associated with cross sections of the 
horizon carrying some {\it punctures}. Spin quantum numbers
$j,m$ attached to the punctures label the quantum states. The 
entropy is obtained by directly counting the possibilities of 
labels that are consistent with a given area
and has been seen to be proportional to the area 
in different approaches \cite{ash,meissner,gm,gm2}.
Numerical counting of the number of states however
has exhibited a modulation of this linear behaviour \cite{cor}.
This is apart from the logarithmic corrections that are also
seen but are well understood \cite{gm0, meissner, gm}.
The breakdown of strict linearity indicates that there is
something beyond the calculations of \cite{ash,meissner,gm,gm2}.
Some attempts have been made
to explain this departure from linearity \cite{sahlmann,agullo}.
But a fundamental issue is involved in connection with
the area. Whereas the earlier
predictions of a linear rise in the entropy were made by
keeping the area only approximately fixed, the numerical
studies fixed the area very precisely. 
This necessitates a fresh consideration of the area constraint. 
The result is an entropy that depends on the
area in a more complicated way where it is {\it bounded}
by a linear rise and saturates the bound at isolated points.

In the analyses of \cite{meissner,gm,gm2}, where the area was only approximately
fixed, the {\it irrational} nature of the quantities $\sqrt{j(j+1)}$ appearing in
the eigenvalues of the area operator was ignored:
essentially, $j$, which is a half-integer or integer, was taken for convenience
as just a real variable. In the present calculation, where the area is precisely fixed, 
the irrational nature of each $\sqrt{j(j+1)}$ must be taken into account. This 
necessitates a classification of spins into isolated classes of compatible spins as 
explained below. Spins of each class are in a kind of thermal contact and equilibrium:
the distribution of the total area amomg different spins is governed
by analogues of temperatures, one for each class. The effective analogue temperature 
depends on the ratios of the contributions to the area of the different classes
of spins. Its variation can explain the oscillations in the variation 
of the number of states with the area.

We first do the counting treating the quantum numbers $j$ as well as the $m$ as labels 
characterizing quantum states, as in \cite{gm}. This is our preferred count,
because the punctures on the horizon are assigned both quantum numbers
$j$ and $m$ \cite{gm}. Thereafter we count the number of states labelled 
by only the $m$, which is a popular counting criterion relying only on the
surface Hilbert space \cite{meissner,gm2}. We end with various concluding remarks.

\section{Counting of states labelled by quantum numbers $j,m$}
It is convenient to use units such that $4\pi\gamma\ell_P^2=1$, where $\gamma$ 
is the so-called Barbero-Immirzi parameter involved in the quantization
and $\ell_P$ the Planck length. Setting the classical area $A$ of the
horizon equal to an eigenvalue of the area operator for a specific spin 
configuration of punctures on the horizon, we write
\beq A=2\sum_{j,m}s_{j,m}\sqrt{j(j+1)},\label{areacon2}\eeq
where $s_{j,m}$ is the number of punctures carrying spin quantum numbers $j,m$. 
Such a spin configuration is admissible if it obeys
(\ref{areacon2}) together with the {\em spin projection constraint}
\beq 0=\sum_{j,m}ms_{j,m},\label{newc}\eeq
which is required for a quantum horizon with the topology of a 2-sphere;
it arises from the consistency of a bulk Hilbert space and a surface
Hilbert space in the theory \cite{ash}.
The total number of quantum states for these configurations
is
\beq d_{s_{j,m}}={(\sum_{j,m} s_{j,m})!\over\prod_{j,m}s_{j,m}!}\;.\label{newco}
\eeq

To obtain the dominant permissible configuration that contributes the largest 
number of quantum states, one may maximize $\ln d_{s_{j,m}}$ by varying 
$s_{j,m}$ subject to the constraints. In \cite{gm}, the area constraint and the
spin projection were considered. Now we have to understand that the strict area
constraint is a very severe one and may be decomposed into {\em several} 
constraints. This is because in varying the states,
we can vary the $s_{j,m}$ only integrally and a sum of these integers multiplied
by the irrational factors $\sqrt{j(j+1)}$ has to be kept unchanged. Not all 
$j$-s can mix with one another in such a variation. The quantities 
$\sqrt{j(j+1)}$ for different $j$ are in general relatively irrational, {\it 
i.e.,} have irrational ratios, {\em except in special cases}. For instance, the
quantities for $j=\frac12$, $j=3$  $j=\frac{25}{2}$ and $j=48$
are in the ratio 1:4:15:56; again, the quantities for $j=1$, $j=8$ and $j=49$ 
are in the ratio 1:6:35, the quantities for $j=\frac32$ and  $j=15$ are in the
ratio 1:8 and the quantities for $j=2$ and $j=24$ are in the ratio 1:10. Thus 
the set of values of $j$ gets divided into disjoint 
subsets such that the $j$ in each subset are compatible with one another
in the sense that the quantities $\sqrt{j(j+1)}$ are
in rational ratios within a subset but in irrational ratios in different subsets. 
When we vary the $s_{j,m}$, mixing can occur only within such a subset but not 
across subsets. So the contribution of each subset to
the total area must remain fixed:
\beq A_N\equiv2\sum_{j\in N,m}s_{j,m}\sqrt{j(j+1)}={\rm const},\eeq
where $N$ denotes such a subset.

Now one can write 
\beq
\delta\ln d_{s_{j,m}}=(\sum_{j,m}\delta s_{j,m})\ln \sum_{j,m}s_{j,m}
-\sum_{j,m}(\delta s_{j,m}\ln s_{j,m})
\eeq
for small $\delta s_{j,m}$.
The condition for the maximum can be expressed in terms of Lagrange multipliers
$\lambda_N,\alpha$:
\beq \ln {s_{j,m}\over\sum s_{j,m}}=-2\lambda_N\sqrt{j(j+1)}-\alpha m, 
({\rm ~for~ }j\in N)\eeq
whence
\beq {s_{j,m}\over\sum s_{j,m}}=e^{-2\lambda_N\sqrt{j(j+1)}-\alpha
m}\quad ({\rm for~ }j\in N).\label{news}\eeq
Consistency requires that $\lambda_N$ and $\alpha$ be related to each other by 
\beq\sum_N\sum_{j\in N}
e^{-2\lambda_N\sqrt{j(j+1)}}\sum_me^{-\alpha m}=1.\eeq 
In order that (\ref{news}) satisfies
the spin projection constraint, we need 
$\sum_mme^{-\alpha m}=0$ for each $j$, which essentially implies $\alpha=0$. 
Therefore, the consistency condition becomes
\beq\sum_N\sum_{j\in N,m}e^{-2\lambda_N\sqrt{j(j+1)}}=1.\label{lambda}\eeq
Here the sum over $N$ goes over the subsets appropriate for the area 
in question.

This is similar to the equation obtained earlier using a single area constraint.
However, while that equation could be solved for its single $\lambda$, now there
are several variables $\lambda_N$ in general. There are also more equations:
\beq A_N=2(\sum s_{j,m})\sum_{j\in N}\sqrt{j(j+1)}(2j+1)
e^{-2\lambda_N\sqrt{j(j+1)}},\eeq
where the $2j+1$ comes from summing over $m$. This implies
\beq {A_N\over A}={\sum_{j\in N}\sqrt{j(j+1)}(2j+1)e^{-2\lambda_N\sqrt{j(j+1)}}
\over \sum_N\sum_{j\in N}\sqrt{j(j+1)}(2j+1)e^{-2\lambda_N\sqrt{j(j+1)}}},\eeq
which, together with (\ref{lambda}), determine the  $\lambda_N$.

Note that  in Stirling's approximation,
\beq \ln d_{s_{j,m}}=\sum_N\lambda_N A_N+\alpha \sum s_{j,m}m,\eeq
in which the last term vanishes,
so that the leading contribution to the entropy is
\beq S=\sum_N\lambda_N {A_N\over 4\pi\gamma\ell_P^2}\eeq
if normal units are used.

It has to be understood that the $A_N$ are determined by $A$ and the
$\lambda_N$ are in turn determined by these. 
It is not possible to solve the equations explicitly in general, but let us 
consider some simple cases. The simplest possible case involves a single subset
$N$ with $j=\frac12, j=3,  j=\frac{25}{2}, j=48,...$.
This corresponds to the area being an integral multiple of $\sqrt{3}$ in our 
units. Numerical solution of (\ref{lambda}) for this special case yields 
\beq\lambda_1^{(j,m)}=0.521...,\eeq
which is greater than ${\ln 2\over\sqrt 3}\approx 0.400...$
corresponding to just spin 1/2 \cite{ash},
but is less than the 0.86 corresponding to the use of a single area
constraint \cite{gm}.

Next we take two $N$-s, one with $j=\frac12, j=3,  j=\frac{25}{2}, j=48,...$
and the other with $j=1, j=8,  j=49,...$, but keep only the
smallest values of $j$ in each subset as an approximation. 
This means that the area must be the sum of an integral multiple
of $\sqrt{3}$ and an integral multiple of $2\sqrt{2}$ in our units.
Then 
\beq A_1&\propto& \sqrt{3}\exp (-\lambda_1\sqrt{3})\nonumber\\
A_2&\propto& 3\sqrt{2}\exp (-2\lambda_2\sqrt{2}).
\eeq
These equations, together with
\beq
2e^{-\lambda_1\sqrt{3}}+3e^{-2\lambda_2\sqrt{2}}=1,\eeq 
determine $\lambda_1,\lambda_2$ as functions of $A_1/A_2$.
The entropy, involving these $\lambda$-s, can
be written as $S=\bar\lambda A$ with an average $\bar\lambda$.
We find that as $A_1/A_2$ varies, $\bar\lambda$ increases from 
$\frac{\ln 3}{2\sqrt{2}}$,
and falls to $\frac{\ln 2}{\sqrt{3}}$ after reaching 
\beq
\bar\lambda|max=0.704 \quad {\rm at ~}A_1/A_2=0.884,\eeq 
corresponding to $\lambda_1=\lambda_2$.

If several $N$ are involved, finding the corresponding
$\lambda_N$ is on the same lines, but more complicated. 
The average $\bar\lambda$ again varies with the ratios of the $A_N$-s and
can be seen to be  maximum when all $\lambda_N$ are equal.
The value of this maximum depends on the subsets $N$ involved, and reaches {\it its}
peak value only when all spins participate; then it becomes
equal to 0.86 by virtue of (\ref{lambda}). Thus $\bar\lambda$ is
{\it bounded} by the value 0.86 corresponding to a single $\lambda$.

In general, as the area $A$
is varied, the relevant subsets $N$ change and so do the
$\lambda_N$.  For values of $A$ which are not sums of eigenvalues,
there is no state at all, but when $A$ can be expressed
as a sum of terms of the form $\sqrt{j(j+1)}$, some
$A_N$s are nonzero. The number of states involves also
the corresponding $\lambda_N$. The quantities $A_N,\lambda_N$ 
fluctuate with variations of $A$ and so does $\bar\lambda$,
which is not a constant but depends on the ratios of the
`components' $A_N$ of $A$.
That is why the variation of the number of states is not monotonic
and a structure is seen. As $\bar\lambda\le 0.86$, the plot is generally
below the $S=0.86A$ line, but touches it at points where all spins 
participate and moreover all $\lambda_N$ are equal, which happens for
special ratios of the $A_N$-s. $\bar\lambda$ falls where the area
is not an eigenvalue. Such oscillations with a linear bound
have been seen numerically \cite{cor}. A step-like structure emerges
on bunching of the area variable into bins because of the discrete
nature of area eigenvalues. 
 
\section{Counting of states labelled by quantum numbers $m$}

The ideas of the previous calculation can be easily extended to do
a fine-grained counting of states labelled by only the
$m$ quantum numbers as envisaged in \cite{ash}.
Here one has to consider $s_m\equiv\sum_j s_{j,m}$ and
maximize the combinatorial factor by varying the $s_{j,m}$.
Maximization of the combinatorial factor occurs not
in the interior but on the boundary of the configuration space \cite{gm2}.
The consistency condition for the Lagrange multipliers $\lambda_N$ becomes
\beq\sum_N\sum_{j\in N}(2+\delta_{j1})e^{-2\lambda_N\sqrt{j(j+1)}}=1,\eeq
where for each $m$ one needs the minimum $j$ possible. Furthermore,
\beq {A_N\over A}={\sum_{j\in N}\sqrt{j(j+1)}(2+\delta_{j1})
e^{-2\lambda_N\sqrt{j(j+1)}}\over \sum_N\sum_{j\in N}\sqrt{j(j+1)}(2
+\delta_{j1})e^{-2\lambda_N\sqrt{j(j+1)}}},\eeq
which, together with the preceding equation, determine the $\lambda_N$.

If one has an $A$ such that only one $A_N$ is nonzero, corresponding to the
subset with $j=\frac12, j=3,  j=\frac{25}{2}, j=48,...$,
the corresponding $\lambda_N$ can be easily determined from the consistency
condition to be 
\beq\lambda_1^{(m)}=0.453...\eeq
This is less than both the value 0.521 obtained above
with $j,m$ counting and the value with a single constraint but $m$ counting,
namely 0.790 \cite{gm2}. 
The case of two nonzero $A_N$ with $j=\frac12..$ and $j=1..$ is the {\it same} 
as in the counting with $j,m$ labels and again produces a peak $\bar\lambda$ 
value of {\bf 0.704}. When there are several nonzero $A_N$ the number of states
is determined by the corresponding $\lambda_N$ which can in principle
be obtained from the above equations. The $A_N,\lambda_N$ fluctuate with change 
of $A$ and produce a structure as in the previous situation, 
with $\bar\lambda\le 0.79$.

\section{Conclusion}

The earlier approximate analytical calculations, 
valid for large areas, indicated that the entropy, 
increases linearly with the area of the horizon up to
a logarithmic correction. 
In this Letter a more precise analytical 
calculation has been carried out, leading to an understanding
of a numerically observed modulation of the linearity. The point
is that when one talks about the area eigenvalues, these involve
irrational numbers $\sqrt{j(j+1)}$. Real numbers in general cannot be
expressed as a combination of numbers of this sort, but some can be.
Such numbers can often be expressed in {\it many different ways}
as combinations of this form. The building units $\sqrt{j(j+1)}$
fall naturally into disjoint classes. Members of a single class
have rational ratios, so that different combinations of members
in a class can be constructed with a constant sum. 
Members of different classes however have irrational ratios.
If the total area is to
be kept constant, the contribution from each of these classes
has to be kept constant separately. Thus there is a number
of constraints corresponding to the different classes. As the
number of punctures is finite for finite $A$, the number of
subsets or constraints also must be finite. In practice
the effective number will be small because only low spins $j$ can
contribute significantly to the counting of states. In any case, each class 
or each constraint is associated with a Lagrange multiplier
which can be determined if the contributions of the different classes
to the total area are known. The number of states can be expressed as
a function of all these variables: the area contributions from the different 
classes and the Lagrange multipliers. While they depend on and are determined
by the total area, they fluctuate as the total area changes: indeed the
contributions of the different classes to the total area change wildly as
different classes of $\sqrt{j(j+1)}$ are involved. On top of this there are
values of the area which cannot be expressed as combinations of $\sqrt{j(j+1)}$
and so yield no quantum state. These fluctuations result in
a non-monotonic change of the number of states with the total area
bounded by a linear growth. These oscillations go over into steps on
consideration of bands of area.

In the above discussion we have considered only the leading area dependence,
ignoring the logarithmic corrections found in earlier papers. Clearly, the
introduction of new constraints has a r\^{o}le to play in the logarithmic 
corrections. The usual term ($-\frac12\ln A$) arises because of the presence
of two constraints, {\it viz.,} the area and spin projection constraints.
Each additional constraint will contribute a $-\frac12\ln A$. However,
the number of new constraints is finite for finite area, so the number
of subsets $N$ is only finite. It is only of academic interest to calculate
$N$ and the logarithmic corrections for a general case.

The above discussion may cause some confusion about the correct value 
of the $\lambda$ or of the Barbero-Immirzi parameter. 
It has to noted that we have calculated the degeneracy of a fixed area eigenvalue.
In contrast, in earlier work, one calculated the number of area 
eigenstates whose corresponding area eigenvalues fell within a narrow band 
$[A-\epsilon,A+\epsilon]$, 
where $\epsilon/A\ll 1$, around a fixed area $A$. When there is a band, 
the area constraint is implemented only approximately: one can always find 
a rational number close to $\sqrt{j(j+1)}$ within the band
and the additional $A_N$ constraints do not arise. Therefore the earlier
calculations were appropriate in a statistical sense. They were
realistic in the sense that the larger an area is, the larger are the errors in fixing
it, hence a band is more appropriate. For a fixed eigenvalue the
number of degenerate states must be smaller than the total number of states
calculated for a band. This is related to the increase in the number of 
constraints in the present calculation. So the smaller and varying values of the
parameters obtained in this paper are not unexpected. The earlier values 
are reached as the
peak values of $\bar\lambda$ as it varies with varying ratios of the $A_N$s.

Each $\lambda$ behaves like an inverse temperature -- but
it is dual to the area rather than to the energy, in the sense that
$\lambda\sim{\partial S\over\partial A}$. The appearance of several $\lambda_N$
corresponds to a many-temperature system. This means that the system as a
whole is not in what may be thought of as an analogue of thermal contact; it 
has different subsystems each characterized by its own analogue temperature.
This occurs because area may be exchanged between spins in one class, but not
between different classes. Of course, these analogue temperatures are not
related in any way to the actual Hawking temperature.

\end{document}